\documentclass[nofootinbib,prd,preprint]{revtex4-1}

\setcitestyle{super,open={},close={}}

\usepackage{setspace} 
\usepackage{etoolbox}
\makeatletter
\patchcmd{\@footnotetext}
{\setspace@singlespace}{0.8}
{}{}
\makeatother
\usepackage{amsmath}
\usepackage{amsfonts}
\usepackage{graphicx}
\usepackage{bm}
\usepackage{subfigure}
\usepackage[dvipsnames]{xcolor}
\definecolor{ReflexBlue}{rgb}{ .0902,.0902,.5882}
\usepackage[breaklinks]{hyperref}
\hypersetup{
    colorlinks=true,
    linkcolor=ReflexBlue,
    filecolor=magenta,      
    urlcolor=ReflexBlue,
    citecolor=ReflexBlue,
}
\begin{document}

\title{How physics got its right hand:\\The origins of chiral conventions in electromagnetism}

\author{Tyler McMaken}
\email{tcmcmaken@umary.edu}
\affiliation{Department of Mathematics and Physics, University of Mary, Bismarck, ND 58504}

\date{\today}

\begin{abstract}
Why do physicists almost universally take the direction of positive rotation to be counterclockwise, and three-dimensional coordinates to be right-handed? This paper traces the historical development of these chiral conventions, with an emphasis on the physical quantity whose direction became the focal point of this discussion in the mid-1800s, the magnetic field. Though these standards are often reduced to mere mathematical, inconsequential choices, an analysis of the impact of Newton, Maxwell, the London Mathematical Society, and others toward the subject can enhance classroom discussion, not only as a contextual sidebar, but also by emphasizing the influence conventions in physics can have on pedagogy, communication, and scientific advancement.
\end{abstract}
% AJP requires an abstract for all regular article submissions.
% Abstracts are optional for submissions to the "Notes and Discussions" section.

\maketitle

\section{Introduction}\label{sec:int}

In both mechanics and electromagnetism courses, a common question that often arises from students is related to the origin of right-handed coordinates and other similar conventions in physics: Why is counterclockwise rotation defined as positive? Who decided that magnetic fields and other axial vectors follow a right-hand rule and not a left-hand rule? Is there a physical reason behind it all, or at least a pedagogical or historical one?

The answer usually given to these questions in textbooks and by many instructors is to brush them off as unimportant\textemdash they are merely conventions and have no bearing on physical observations, so one should take care only to be consistent and acquainted with the greater scientific community's practices. Any further consideration would then be unproductive and take time away from the issues that \emph{really} matter. Some go further to give speculative or even erroneous explanations for the origin of these conventions, like that right-handed coordinates became conventional because the majority of people are right-handed, or that counterclockwise angles have been the norm among mathematicians since at least the time of Babylonian astronomy.

% or that the first person to fix the convention in its modern form was John Ambrose Fleming at the close of the nineteenth century.\footnote{Indeed, Fleming provides perhaps the earliest published reference to a modern ``right-hand rule'' using the right hand as a mnemonic for the direction of electromagnetic quantities; however, as will be seen, the right-handed convention had been fixed in physics several decades earlier, the term ``right-handed'' had been used to describe the convention at least since the time of the Industrial Revolution (first in the context of screws), and more general bodily mnemonics for the convention had existed since the earliest days of electromagnetism in the 1820s.}

Readers may be surprised to learn that the aforementioned conventions were only fixed because of a vote of the London Mathematical Society in 1871 at the behest of James Clerk Maxwell. The goal of this work is to recount the fascinating story of the historical development of chiral conventions in physics, culminating in Maxwell's efforts to understand and create a shared language for the axial vectors of electromagnetism.

To help frame the discussion, consider Faraday's law of induction, which states that an electromotive force $\mathcal{E}$ around a conducting loop is caused by a change in magnetic flux through a surface $\Sigma$ enclosed by that loop (for example, when a bar magnet is moved toward a loop of wire to induce a current):

\begin{equation}\label{eq:faraday}
    \mathcal{E}=-\frac{d}{dt}\iint_\Sigma\bm{B}\cdot d\bm{A}.
\end{equation}

It is a useful exercise for students to reflect on how changes in underlying conventions would affect a given equation. Some conventions, like the choice of area vector $d\bm{A}$ as into or out of the loop, are completely arbitrary and must be decided for each given setup, while others, like the three outlined below, have been fixed in advance by scientific consensus. In particular, consider what would happen to the above form of Faraday's law if (1) the sign of electric charge (and therefore the direction of conventional current and the sign of $\mathcal{E}=dW/dq$) were reversed, (2) the north/south polarity of magnetism (and therefore the conventional direction of $\bm{B}$-field lines) were reversed, or (3) the right-hand rule were exchanged for a left-hand rule. In all three cases, the minus sign of Eq.~\ref{eq:faraday} would become a plus sign.

Nonetheless, for some problems in physics (\textit{e.g.}, Gauss's law), the equations are completely unaffected by the three above conventions. Often, this is the result of an internal cancellation of two sign changes; for example, students switching to a left-hand rule when using Eq.~\ref{eq:faraday} will still obtain the same direction for the induced current if they also use a left-hand rule for Amp\`ere's law when determining the direction of the magnetic field from a source current. In this case, it becomes apparent that Lenz's law is not encoded in the minus sign of Eq.~\ref{eq:faraday}. Even if the magnetic field direction were reversed such that $\mathcal{E}=+d\Phi_B/dt$, students can check that the induced magnetic field still opposes the original field, as a result of the relative sign difference between the two dynamical Maxwell equations (and ultimately from Lorentz covariance of the electromagnetic field tensor).

The above enumeration of conventions baked into Eq.~\ref{eq:faraday} will serve as an outline for the remainder of this paper as we proceed chronologically through history. To begin, Sec.~\ref{sec:CCW} will discuss a mathematical convention that, like $d\bm{A}$, is completely arbitrary (but, as will be seen, had a profound impact on the later right-hand rule): choosing the positive angle in polar coordinates to be counterclockwise. Then, Sec.~\ref{sec:Q} will discuss the sign of electric charge, as defined by Benjamin Franklin, and Sec.~\ref{sec:mag} will discuss the polarity of magnetism, as defined by Faraday and other early nineteenth century figures. Finally, Sec.~\ref{sec:LMS} will discuss how all these conventions were solidified and supplemented by the decision to follow right-handed coordinates in the time of Maxwell.

% Alongside this discussion, in order to showcase the diversity of mathematical conventions used by past physicists, the supplementary material presents in modern notation the dozen coordinate systems used by Newton in his \textit{Method of Fluxions}, which contains perhaps the earliest recorded example of polar coordinates.
Alongside this discussion, in order to showcase the diversity of mathematical conventions used by past physicists, the Appendix presents in modern notation the dozen coordinate systems used by Newton in his \textit{Method of Fluxions}, which contains perhaps the earliest recorded example of polar coordinates.

Finally, it should be noted that this work is not intended solely to give a history lesson. While historical developments in physics are important topics for the classroom to contextualize and personalize scientific learning, this particular story also provides important insights about the role of definitions in physics, the importance of linguistic and mathematical consensus in communication, and the way in which students can balance old and new conventions in their own work. A discussion furthering these points will conclude the work in Sec.~\ref{sec:dis}.

\section{Proto-right-handed conventions: Polar coordinates}\label{sec:CCW}

To begin, one of the primary reasons cited by the London Mathematical Society (as will be recounted in Sec.~\ref{sec:LMS}) for the adoption of a right-handed rather than left-handed system had to do with how mathematicians of the day were already used to orienting two-dimensional coordinates on a black board. The choice to draw the positive $x$-axis to the right may be inferred as a natural byproduct of left-to-right written language conventions, but one may wonder how the choice the draw polar angles counterclockwise from this axis first came about.

A review of the historical data reveals that no rotational conventions were standardized among mathematicians up to and including the inventors of polar coordinates themselves, though if anything, clockwise angles were preferred. Early works requiring the definition of an angle were focused on mathematical spirals, following Archimedes' influential work \textit{On Spirals} around 225 BC.\cite{Archimedes225} Despite how they are usually presented today, Archimedes drew all his namesake spirals clockwise outward, and likewise, geometers of the seventeenth and eighteenth centuries like Cavalieri\cite{Cavalieri1635} and Fontana\cite{Fontana1780} used clockwise angles to solve similar types of problems. Priority in the publication of polar coordinates is traditionally given to Jakob Bernoulli, who in the January 1691 edition of \textit{Acta Eruditorum} analyzed points of a spiral in terms of a radial line and the distance along a circular arc.\cite{Bernoulli1691} This angular distance started from the top of the vertical axis and extended clockwise downward.

However, Bernoulli had no reason to standardize this methodology across his works, and so he and his successors used a potpourri of both clockwise and counterclockwise conventions throughout the next two centuries. Euler in 1748 included examples both with counterclockwise angles from the positive $x$-axis and clockwise angles from the negative $x$-axis,\cite{Euler1748} and Wessel in 1797 defined positive as ``with the Sun'' (\textit{i.e.,} clockwise) in the first work to give a geometric interpretation to Euler's formula as rotation in the complex plane,\cite{Wessel1797} while a similar yet more widely popular work by Argand in 1806 displayed all figures with counterclockwise angles.\cite{Argand1806} A decade later, the first work to use the English term ``polar coordinates''\cite{Lacroix1816} likewise portrayed many rotations as counterclockwise, but it did not commit to a universal convention. Indeed, mathematicians saw the importance of being comfortable with any coordinate definition suited to the problem at hand.

\begin{figure}[t!]
\centering
\includegraphics[width=0.35\textwidth]{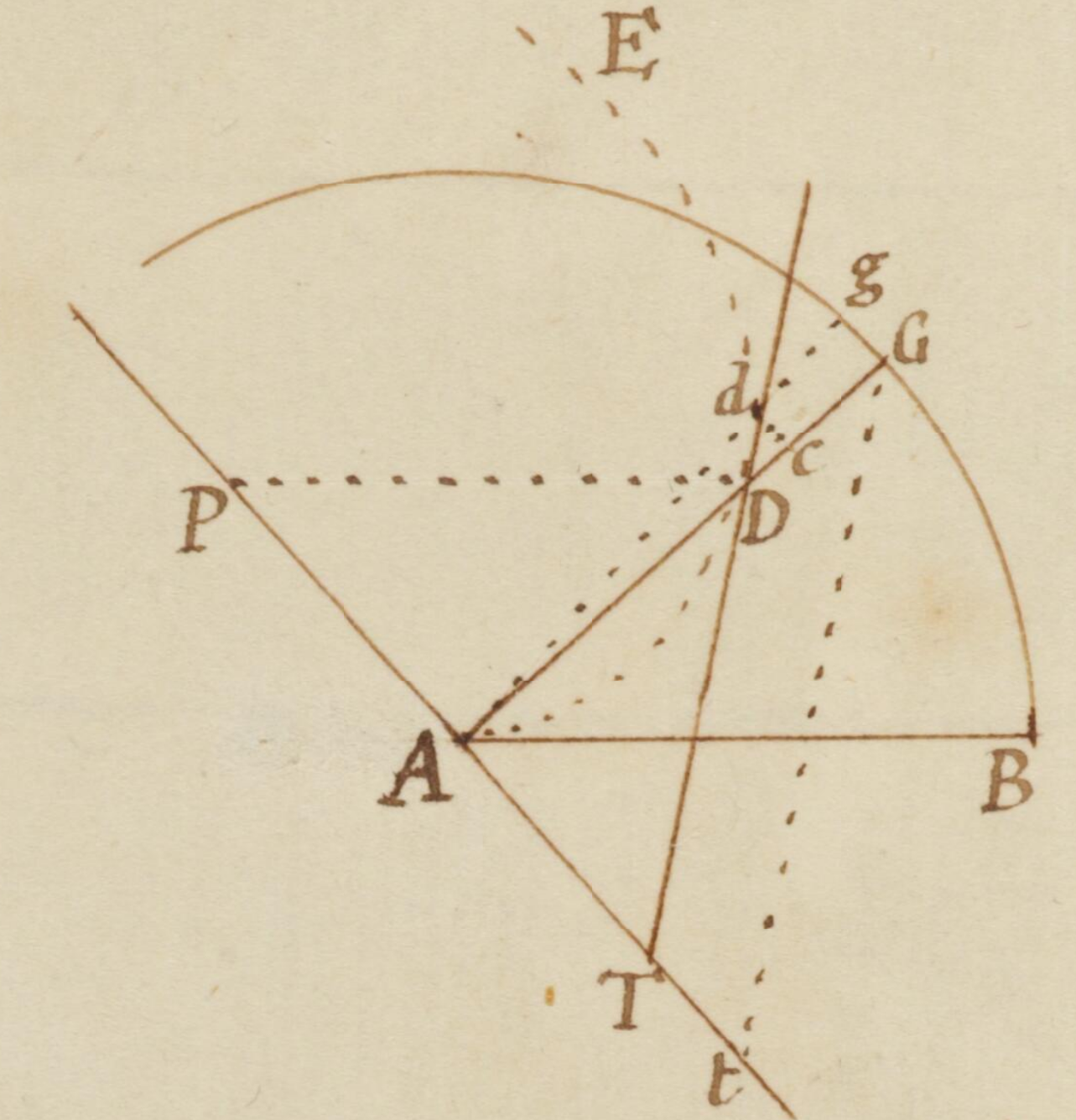}
\caption{Figure from the section ``Seventh Manner: For Spirals'' from Newton's handwritten draft of \textit{Method of Fluxions}, which provides one of the earliest uses of polar-type coordinates. To find the tangent DT of the dashed spiral ADE at point D, Newton chooses the $x$ coordinate to be the counterclockwise circumferential length BG and the $y$ coordinate to be the radial length AD. (Cambridge University Library, MS Add.\ 3960, p.\ 3:25)}
\label{fig:newton}
\end{figure}

Perhaps the most important yet overlooked source of polar coordinates that led to the increased popularity of counterclockwise angles is Sir Isaac Newton, who in his \textit{Method of Fluxions} (written 1671 but published posthumously in 1736) supplied nine (or ten or twelve, depending on how they are grouped) different coordinate systems with which to do calculus.\cite{Newton1736} His ``seventh manner,'' shown in Fig.~\ref{fig:newton}, with an angle extending counterclockwise from a rightward horizontal line, provides the earliest written use of a polar coordinate system in fully analytic form, with a setup that is surprisingly similar to the modern convention.\cite{Boyer1949} Perhaps Newton's work reveals the early customs that led to those of the London mathematicians two centuries later.

%Since scant attention has been paid in the literature to Newton's coordinate systems in \textit{Method of Fluxions}, these coordinates are enumerated in modern notation in the supplementary material. This enumeration is interesting because it shows the diversity of conventions that physicists have historically used to solve problems beyond today's standard Cartesian and polar coordinates.
Since scant attention has been paid in the literature to Newton's coordinate systems in \textit{Method of Fluxions}, these coordinates are enumerated in modern notation in the Appendix. This enumeration is interesting because it shows the diversity of conventions that physicists have historically used to solve problems beyond today's standard Cartesian and polar coordinates.

\section{How was the sign of electric charge decided?}\label{sec:Q}

Following the death of Newton, the next generation of scientists began to realize the importance of applying mathematical conventions to physical phenomena, especially those involving electricity and magnetism. This story of conventions had an unfortunate yet pivotal start in 1751 with Benjamin Franklin's definition of positive and negative charge. Prior to that point, static electricity created in Leyden jars was hypothesized to exist as two separate fluids, a ``vitreous'' fluid created from rubbing \textit{vitrum} (\emph{i.e.}, glass) with silk, and a ``resinous'' fluid created from rubbing resins like amber with fur. Franklin's groundbreaking proposal was that these two fluids were actually positive and negative portions of the same ``electrical fire.''\cite{Franklin1751}

This new way of discussing the phenomenon laid the foundational framework necessary for later electrodynamic discoveries like galvanic cells and the existence of charge carriers, with an unfortunate consequence: the direction of conventional current was set as traveling from a source with positive charge toward a negative charge, but it was later found that the carriers of charge in Leyden jars were sourced from the resinous fluid, which Franklin had called negative. To this day, to the dismay of generations of frustrated students, physicists have stuck with the backwards convention that current flows in the opposite direction as the electrons that constitute the flow in usual solid conductors.

Calling Franklin's electric charge sign convention ``backwards'' is justified pedagogically by the fact that the convention is tied to a physical observable, the flow direction of mobile charge carriers. Though there are exceptions (\textit{e.g.}, electron holes in semiconductors, or positive ions in an electrolyte), the charge carriers for the vast majority of electrodynamic systems are negatively-charged electrons. Given the fairly ubiquitous notion that ``positive'' implies a surplus of something, conventional current would most straightforwardly assign a positive sense to the most common constituents of electric flow, just as one would when defining positive pressure in fluid flow or positive temperature in heat flow.

However, a natural choice is not readily apparent for the sign convention of the next phenomenon to be analyzed, the magnetic field. Classically,\footnote{The qualifier "classically" is a crucial one, since at a quantum mechanical level, it has been known since the 1950s that the direction of a particle's intrinsic magnetic moment can be disambiguated by observing its interactions with the weak nuclear force, which only couples to what the modern convention calls left-handed particles (\textit{i.e.}, the conventional direction of spin is always in the opposite direction as the particle's momentum, in the massless limit). In this narrow sense, the magnetic field direction is just as backwards as the direction of conventional current.} there is no unambiguous physical reason to give priority to one of the poles of a magnet over the other as positive or negative, in the same way that the rotational axis of a spinning body like the Earth could equally be defined as pointing north or south. How then did physicists decide which pole should be positive?

\section{Early developments for the polarity of magnetism}\label{sec:mag}

The earliest reference to a magnetic field with a sense of directionality comes from Michael Faraday, the self-taught scientist who originated the concept of magnetic lines of force.\cite{Faraday1831} He was not the first to understand that magnets and current-carrying wires produce forces, but prior laws by Amp\`ere, Biot, and Savart focused solely on the direct interaction between two currents without reference to a mediating field. Faraday finally made the switch from thinking in terms of Amp\`erian elementary current loops to magnetic lines of force in 1831, and he had much difficulty working out the directional conventions for these lines, infamously noting that it was ``very simple, although rather difficult to express.'' In fact, he initially got the direction wrong for his own law of induction in several manuscripts before setting things straight the following year.\cite{Romo1994}

\begin{figure}[b!]
\centering
\includegraphics[width=0.4\textwidth]{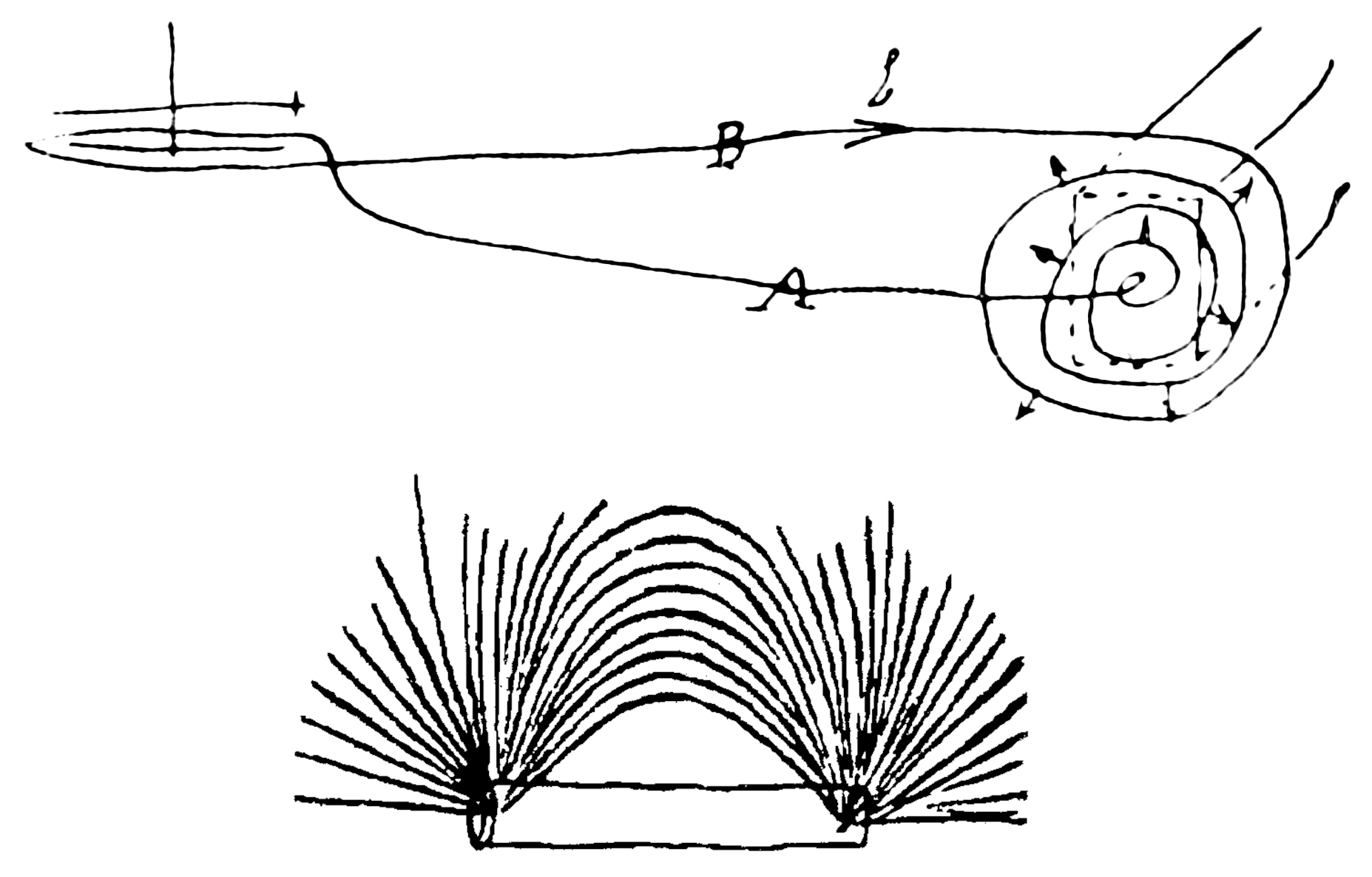}
\caption{Faraday's sketches from his 1831 experiments. (Top) If the spiral of wire shown is pushed into the page toward the marked (north) pole of the magnet, a current of electricity $\mathcal{E}$ will be induced in the direction shown. The small arrows intersecting the spiral, which point out from the marked pole, indicate one of Faraday's earliest attempts to visualize ``magnetic curves.''\cite{Faraday1831} (Bottom) Elsewhere, he depicted these curves without a sense of directionality, showing only which end of the magnet was marked or unmarked.\cite{Romo1994}}
\label{fig:faraday}
\end{figure}

\begin{figure}[t!]
\centering
\includegraphics[trim=8.2cm 12.3cm 7.1cm 7cm,clip,width=0.21\textwidth]{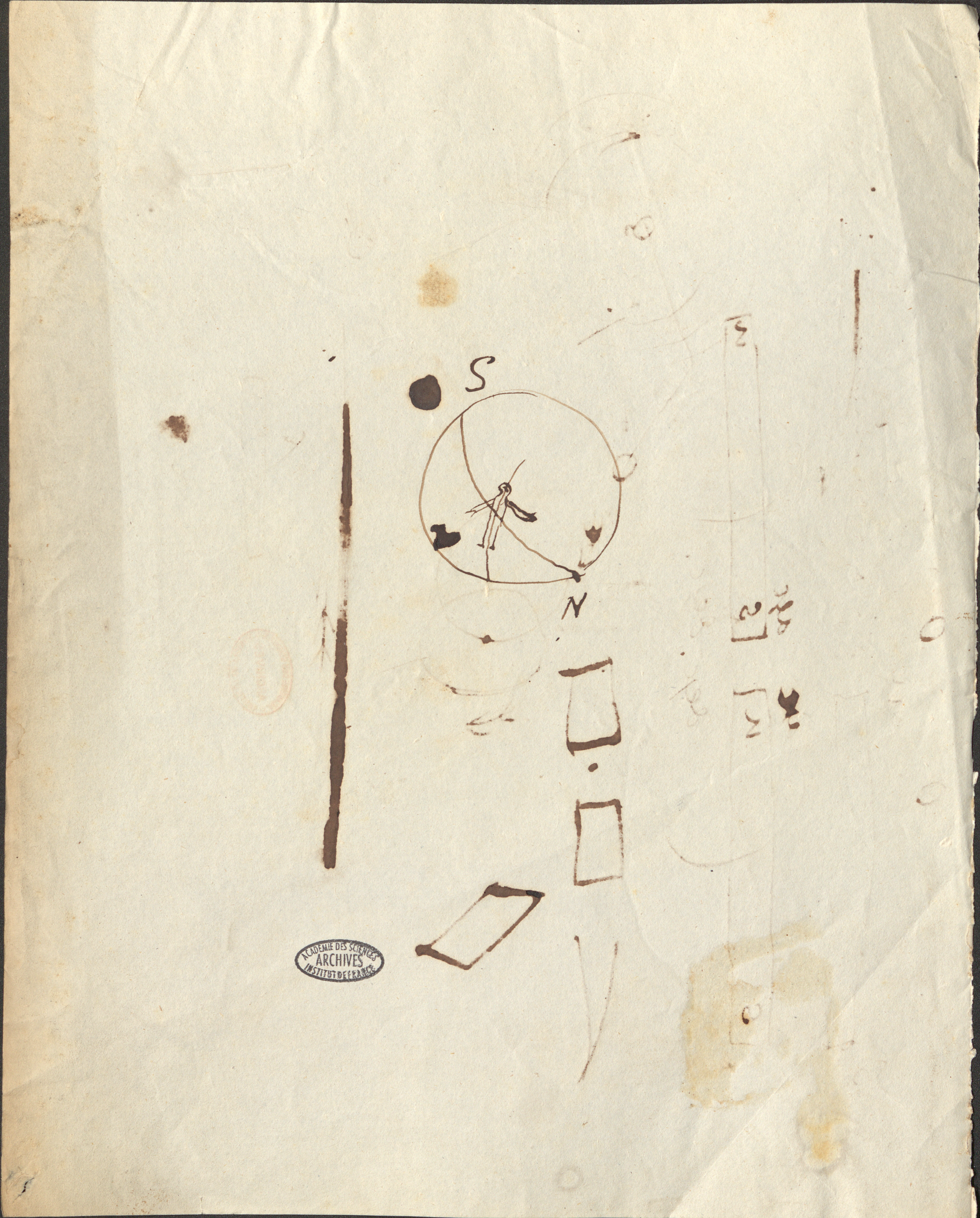}
\caption{Amp\`ere's little man lies on his back along a partially-drawn equator, with a geographic north-south meridian crossing him left-to-right. An electrical current running along the equator from his feet to his head will produce a magnetic field, which will cause a north-pointing compass placed above him to follow the direction of his left arm.\cite{Ampere}}
\label{fig:bonhomme}
\end{figure}

Faraday, who used no mathematics in his research, mostly abstained from indicating any sense of directionality for his ``magnetic curves,'' though he did carefully note any direction of physical motion (\textit{e.g.}, the rotations of his homopolar motor or of polarized light). However, as shown in Fig.~\ref{fig:faraday}, early sketches indicate a preference for curves pointing in the same direction that the north end of a compass would follow. This sense of direction had been implicit in the minds of scientists for the past decade, ever since Hans Christian \O{}rsted in 1820 discovered while setting up for a lecture that a current-carrying wire can deflect a magnetic compass needle from a distance. Soon after, Amp\`ere devised a rule for determining the direction of deflection: a little man placed over a wire with current traveling from his toes to his head will see a compass placed in front of him point toward his outstretched left arm. (Yes, the first right-hand rule was actually a left-arm rule.)\cite{Greenslade1980} This \textit{bonhomme d'Amp\`ere}, shown with Earth's magnetic field in Fig.~\ref{fig:bonhomme}, became the standard for French students throughout the next century, while the English-speaking students used a later mnemonic developed by Maxwell, his namesake corkscrew.

Thus, the direction of magnetic field lines was, at the earliest stages, tied to the orientation of early nineteenth-century European compasses, whose labeled north and south magnetic poles must by the laws of magnetism be opposite polarity to those of Earth's geographic north and south poles. But why do compass needles point north in the first place? In fact, the earliest compasses from China in the fourth century BC pointed south. But once European maritime explorers gained global influence a millennium and a half later, they standardized their compasses to point in the direction of the star they dubbed ``Polaris,'' which over that time frame had conveniently precessed into the vicinity of the previously vacant celestial north pole.\footnote{For example, in the 4th century BC, Guiguzi mentions that the people of Zheng carried a ``south-pointer'' device to help find their way (Ch.\ 10 of his eponymous work). Around the same time in Greece, the sailor Pytheas undertook northern voyages that were later recounted by Hipparchus, finding that ``there is not a single star at the [north celestial] pole, but an empty spot'' (\textit{In Arat.\ et Eud.\ Phaen.} 1.4.1).}

\section{Vines, hops, and the London Mathematical Society}\label{sec:LMS}

Though the story of the magnetic field's direction appears to have reached a tidy conclusion, mathematical reform in the mid-nineteenth century led to an upheaval and complete reexamination of electromagnetic conventions. This reform began in 1843 when the mathematician Sir William Rowan Hamilton introduced his framework of quaternions, the precursor to modern vector analysis.\cite{Hamilton1847} For the first time, directional quantities in physics gained algebraic meaning and necessitated systematization. But in three dimensions, two distinct arrangements of coordinate axes for vectors are possible, and scientists soon became split as to whether the system should be left-handed ($i$ pointed south, $j$ pointed west, and $k$ up to the sky) or right-handed ($i$ pointed south, $j$ pointed west, and $k$ into the ground). The earliest texts on quaternions written by Hamilton\cite{Hamilton1847} (and later ones by Johann Listing\cite{Listing1847} and Peter Tait\cite{Tait1867}) used left-handed coordinates, while another text coauthored by Tait and Sir William Thomson (later Lord Kelvin)\cite{Thomson1867} used right-handed coordinates.

It is also important to note here that the reference to ``hands'' and ``handedness'' at the time primarily held meaning in reference to the threads of screws\textemdash a right-handed screw must be turned clockwise ($i$ rotating into $j$) to travel away from the observer (the direction of $k$), while a left-handed screw must be turned counterclockwise. A separate set of terms for the coordinate systems suggested by the crystallographer William Hallowes Miller was that of vines and hops.\cite{Maxwell1873} He had read from the biologist Carl Linn\ae us that the tendrils of a vine ascended in a right-handed spiral, while the tendrils of a hop plant ascended in a left-handed spiral, and therefore it became common to use those plants to label the systems.\footnote{The terminology is complicated by the fact that a group of botanists as early as 1827 had begun to use the opposite terms, calling hop tendrils ``right-wound'' where Linn\ae us in the 1700s had called them ``left-wound'' (\textit{volubilis sinistrorsum}). Their reasoning was that these tendrils were constantly turning rightward as they climbed, and even though they traveled left-upward when viewed from the outside, it made no sense to define a plant's movements based on an onlooking animal. This way of thinking also entered into those studying circular polarization of light, which has caused two competing conventions to persist to this day.} Evidently the debate had extended beyond mathematics into the realm of preferences between wine and beer.

Amidst these debates over coordinate conventions arose James Clerk Maxwell, the author of \textit{A Treatise on Electricity and Magnetism},\cite{Maxwell1873} which became the standard text on electromagnetic theory for generations to come. When preparing this manuscript, Maxwell decided to follow Hamilton's left-handed system. But with this convention, Maxwell realized that the direction of the magnetic field would be affected. If a loop of wire is placed vertically (like a wall-hanging clock) with a clockwise electric current, then a north-pointing compass placed at the center of that loop will point away from the observer, but the natural third direction in Maxwell's left-handed system would be toward the observer, leading to an additional minus sign in equations like Amp\`ere's law. His solution? Redefine the magnetic field to point the opposite direction as a compass. After all, it only made sense to define Earth's geographic north pole as the positive magnetic pole and to let Earth's (external) field lines point from the Arctic to the Antarctic, ``boreal'' to ``austral,'' positive to negative.

Despite Maxwell's confidence in his left-handed electromagnetism (in a debate with Tait he adamantly wrote, ``If I find a watch I will pin it to the North star and it will indicate which way the world goes round to all the northerners''\cite{Maxwell1990}), his primary concern was to create a unified system that would be accepted by the whole community rather than forge his own path. This desire for uniformity had previously led him to propose standardized namings for the differential operators $\vec{\nabla}\times$, $\vec{\nabla}\cdot$, and $\vec{\nabla}^2$ (which he respectively called curl, convergence, and concentration\cite{Maxwell1869}), and his namesake equations themselves were an attempt to classify and unify half a century of past research on disparate physical phenomena. In regard to the right- versus left-handed systems, since Maxwell was new to the idea of quaternions and was unsettled by the community's variance in conventions, he wrote to Tait in a postcard dated 8 May 1871, ``I am desolated! I am like the Ninevites! Which is my right hand? Am I perverted?\footnote{``Perversion'' here carries a connotation of error in religious belief (continuing his Biblical allusion to the story of Jonah and the Ninevites), but more importantly it was used by the mathematician Listing\cite{Listing1847} to denote reversal of chirality.} a mere man in a mirror, walking in a vain show? What saith the Master of Quaternions? [...] I must get hold of the Math.\ Society and get a consensus on the craft.''\cite{Maxwell1990}

Indeed, that Thursday evening Maxwell did consult the London Mathematical Society at their May meeting. Two mathematicians, Thomas Hirst and Sir William Thomson, both presented arguments in favor of the right-handed system. The first argument was a physical one, and the second mathematical. Physically, assuming north is defined as positive, the rotation of Earth and the orbit of all the planets would yield right-handed motion. And second, it was noted that mathematicians usually drew right-handed coordinate axes on the black board, with $x$ pointing right, $y$ pointing up, and, as Tait commented, ``You do not fancy the third axis to be running away from you? Eh?''\cite{Maxwell1990}\footnote{Indeed, it has been found from modern studies of spatial reasoning and physical awkwardness in problems involving hand mnemonics that choosing a third axis toward rather than away from the observer is pedagogically favorable.\cite{Kustusch2016}}

Curiously enough, no arguments were presented in favor of the left-handed system; Hamilton had passed away five years prior, Listing lived in Germany, and Tait had only used the left-handed system ``so as to avoid perplexing readers of H[amilton].''\cite{Maxwell1990}
Thus, by unanimous consent, the Society concluded in their proceedings:\cite{Maxwell1990}

\begin{quote}
    System of the Vine adopted May 11, 1871, by the L.M.S.\\
    Vitreous electricity is $+$.\\
    Austral magnetism is $+$.\\
    {[$\ldots$]} Magnetization is from the S end to the N end of magnet.
\end{quote}

Maxwell subsequently revised his entire treatise to follow the right-handed convention just in time for final draft submission. With the mathematical standard fixed, the Earth's geographic north pole became a magnetic south pole, counterclockwise rotation became positive in all the sciences, and the first set of mnemonic hand rules utilizing the convention appeared from Sir John Ambrose Fleming a decade or two later.\cite{Fleming1902}

\section{Discussion}\label{sec:dis}

Here it has been seen how the conventions of the right-hand rule, counterclockwise positive rotation, and electric and magnetic polarity were established, in part through the diligence of Maxwell in desiring a uniform, coherent system in which electricity and magnetism could be expressed using vectors. As the London Mathematical Society wrote, ``In pure mathematics little inconvenience is felt from this want of uniformity; but in astronomy, electro-magnetics, and all physical sciences, it is of the greatest importance that one or other system should be specified and persevered in.''\cite{Maxwell1990} Thus underlies the significance of the role of conventions in physics\textemdash though proofs in mathematics can employ and interchange whatever assumptions and definitions are most suitable for each problem at hand, conventions in physics are interconnected and guided by observation. Various fields of physics like mechanics and electrodynamics often overlap in their cohesive descriptions of phenomena, so that if separate conventions were used to define, say, the handedness of torque and that of the magnetic field, the system as a whole would yield inconsistencies when analyzing motors. Individual conventions in physics are therefore not arbitrary but instead have been carefully crafted to provide a coherent description of phenomena, as in the conventions of Faraday's law of induction discussed in Sec.~\ref{sec:int}.

Pedagogically, what can students take away from these stories? First, learning about the origins of magnetic research and right-handed conventions can help reinforce curricular topics like the application of Amp\`ere's law or the proper usage of the right-hand rule in various contexts. Second, historical narratives can help to personalize science, so that students understand physics not as an ethereal set of equations to be memorized and employed but rather as an interconnected web of thought in which they can participate, with no admonition against making mistakes (as Faraday's directional blunders demonstrate). And third, students should not be discouraged from asking why things are the way they are. For beneath every mathematical convention or choice of notation is an opportunity to learn and to unravel a complex web of underlying physical motivations and historical biases. One may then understand that the choices of conventional current (which mismatches electron momentum) and the magnetic moment direction (which mismatches neutrino momentum) may seem to ``go against nature,'' but as long as there is scientific consensus, these conventions cause no trouble except a few additional minus signs (\textit{e.g.}, when calculating Hall voltage in the former case, or in the latter case, having charged currents in the weak-sector Lagrangian with the left-handed form $J^\mu_\pm=\frac12\bar{\Psi}\gamma^\mu(1-\gamma^5)\Psi$ rather than the more straightforward $J^\mu_\pm=\frac12\bar{\Psi}\gamma^\mu(1+\gamma^5)\Psi$).

As Maxwell wrote to the London Mathematical Society when discussing the power of quaternions and the naming of vector operators, ``I think that the progress of science, both in the way of discovery, and in the way of diffusion, would be greatly aided if more attention were paid in a direct way to the classification of quantities.''\cite{Maxwell1869} Such attention was paid by Franklin when he made the simple switch from the terms ``vitreous'' and ``resinous'' to ``positive'' and ``negative'' charge, which had a profound effect on scientific progress in the field. Similarly, Maxwell mentioned the importance in his own day of Hamilton's quaternions in revealing the relationships between scalar and vector quantities. He also mentioned the contemporary debate between Helmholtz, whose hydrodynamic analogy treated magnetic fields like translational vectors and electric fields like rotational (axial) vectors, and Amp\'ere and his followers, who treated magnetism as rotational and electricity as translational. With modern special relativity it is now known that both systems are Hodge duals of one another and form equally coherent theories of electromagnetism,\footnote{In fact, one may realize that right-handed conventions can be avoided altogether if formulating electromagnetism in terms of an antisymmetric rank-2 field tensor. However, even in this context, new conventions inevitably arise, such as the choice of metric signature.} but Maxwell's choice to follow the latter convention helped maintain continuity with past scientific progress and made more fundamental connections to the most pertinent observable quantities. One may wonder what classifications in the modern era could help bring physics to new heights of advancement.

In his treatise on electromagnetism at the turn of the twentieth century, the right-hand-rule pioneer Fleming noted that magnetic directionality rules are ``merely a convention or arrangement, and there is nothing except the convention to fix which pole should be selected as the one from or to which we reckon[...] Just as we familiarly speak of the sun as rising and setting, when the effect is really due to the rotation of the earth, so the ordinary language we use in speaking about electric currents flowing in conductors retains the form impressed upon it by older and erroneous assumptions as to their nature.''\cite{Fleming1902} Whenever students engage with physics, they must reckon with these historical assumptions to understand the origin of the linguistic and mathematical conventions that pervade every formula and diagram. By tasking students with determining which physical formulas would change (and how) as a result of specific conventional choices, instructors can help encourage this critical thinking and embody Maxwell's own attitude, that ``it is a good thing to have two ways of looking at a subject, and to admit that there \textit{are} two ways of looking at it.''\cite{Maxwell1856} The moment students start asking ``why'' is the moment they can begin to find pedagogical clarity and perhaps even discover something new.

%\section{Supplementary Material}\label{sup}
%Please click on this link to access the supplementary material, which includes a modern presentation of Newton's coordinate systems from his \textit{Method of Fluxions}. Print readers can see the supplementary material at [DOI].

\section{Appendix}\label{sup}

This Appendix gives a modern presentation of the coordinate systems used in the nine ``manners'' (Latin \textit{modi}) of Newton's \textit{Method of Fluxions}, which includes one of the earliest recorded uses of polar coordinates (his seventh manner). After demonstrating how to find the derivative of any point along a curve geometrically using his method, Newton writes, ``And from hence (I imagine) it will be sufficiently manifest, by what methods the Tangents of all sorts of Curves are to be drawn. However it may not be foreign from the purpose, if I also shew how the Problem may be perform'd, when the Curves are refer'd to right Lines, after any other manner whatever: So that having the choice of several Methods, the easiest and most simple may always be used.''\cite{Newton1736}

\begin{figure}
    \centering
    \subfigure[]{\includegraphics[width=0.32\textwidth]{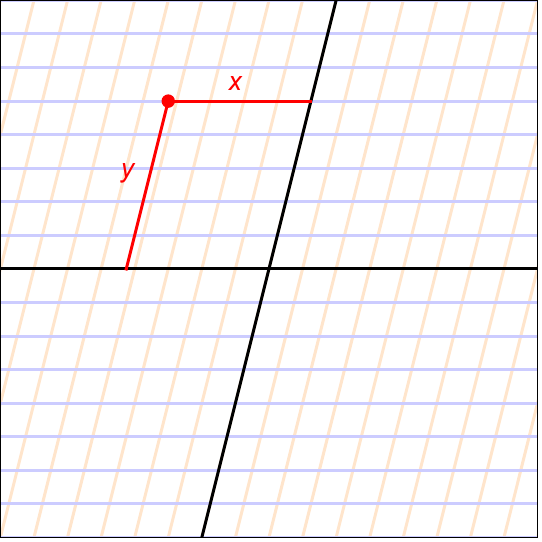}} 
    \subfigure[]{\includegraphics[width=0.32\textwidth]{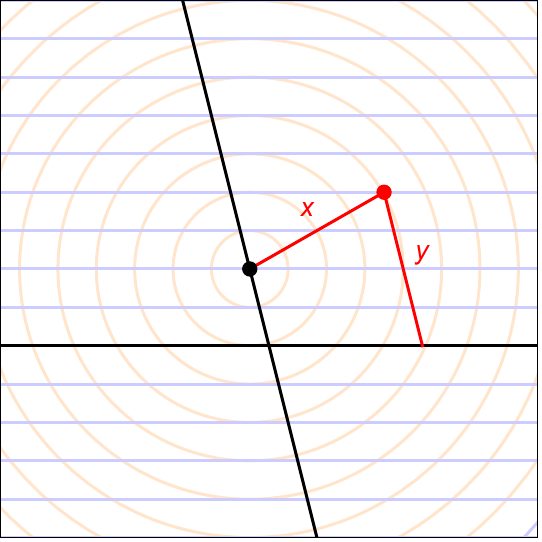}} 
    \subfigure[]{\includegraphics[width=0.32\textwidth]{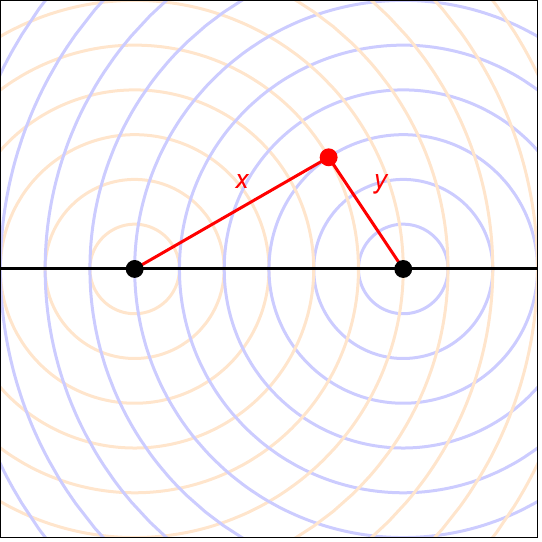}}
    \subfigure[]{\includegraphics[width=0.32\textwidth]{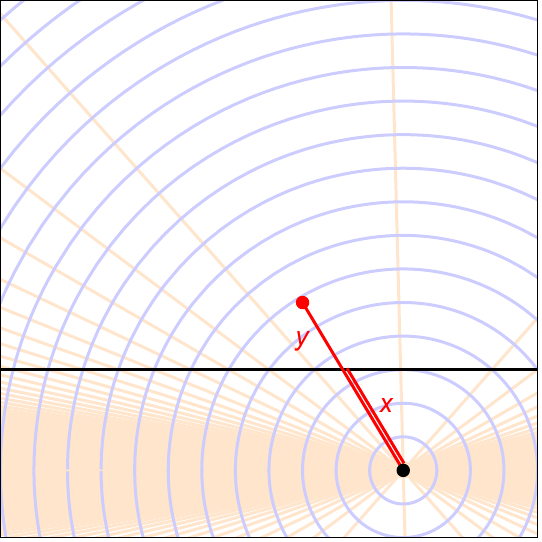}}
    \subfigure[]{\includegraphics[width=0.32\textwidth]{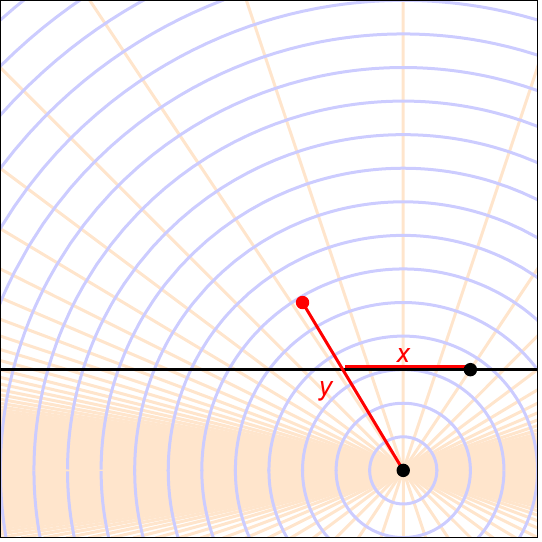}}
    \subfigure[]{\includegraphics[width=0.32\textwidth]{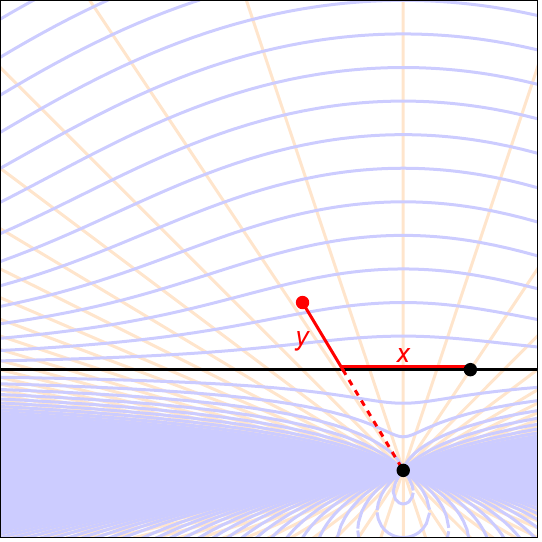}}
    \subfigure[]{\includegraphics[width=0.32\textwidth]{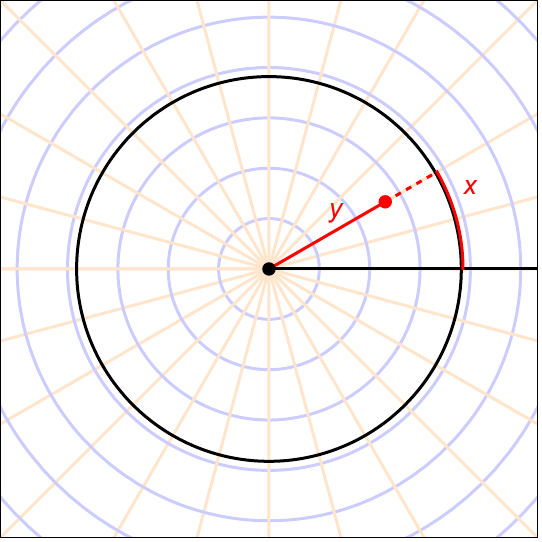}}
    \subfigure[]{\includegraphics[width=0.32\textwidth]{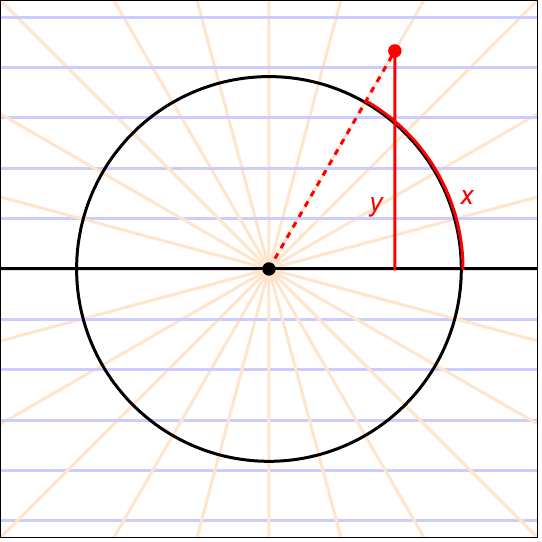}}
    \subfigure[]{\includegraphics[width=0.32\textwidth]{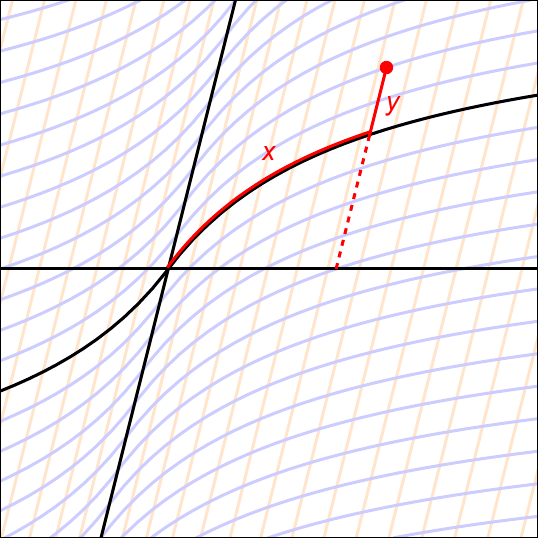}}
    \caption{Coordinate systems (cast in modern form; see explanation in text) used by Newton in each of his nine manners from Problem IV (``To draw Tangents to Curves'') of \textit{Method of Fluxions}. They are labeled in the work as follows: (a) First Manner. (b) Second Manner. (c) Third Manner. (d) Fourth Manner. (e) Fifth Manner. (f) Sixth Manner. (g) Seventh Manner: For Spirals. (h) Eighth Manner: For Quadratrices. (i) Ninth Manner.}
    \label{fig:coord}
\end{figure}

The emphasis in presenting and describing these manners will be on making connections to modern classifications, as a witness to the importance given throughout the main text of clarifying mathematical conventions for proper pedagogy. The coordinates used in each manner are presented in Fig.~\ref{fig:coord}, with unnecessary lines, curves, and labels dropped from Newton's original figures to clarify the underlying coordinate systems. Black lines, curves, and dots in these plots represent the given geometrical aspects of each coordinate system (\textit{e.g.}, axes or radial center points), while the red dot represents a sample point to be specified by the ($x,y$) coordinates shown by the labeled red lines. Contours with $x$ held constant are shown in orange, while contours with $y$ held constant are shown in blue. The spacing between these grid lines is chosen such that the lengths of $x$ and $y$ increase by the same unit amount for each successive contour.

\textbf{First Manner:} The numbering of Newton's coordinate systems is muddled by the fact that in his first manner, he uses examples with two different coordinate systems by modern standards, orthogonal Cartesian coordinates and skew Cartesian coordinates. The more general of the two is presented in Fig.~\ref{fig:coord}a, with $x$ extending horizontally from the non-horizontal axis and $y$ extending from the horizontal axis at a given angle. The curve $y=ax$ in this system is a straight line.

\textbf{Second Manner:} $x$ extends radially from a given point, while $y$ is the same skew Cartesian ordinate as in the first manner. This coordinate system has no straightforward modern classification, but it might be called (along with the eighth manner) ``semi-polar.'' In this and most other manners involving a radial coordinate, the system is not truly a bijective covering of $\mathbb{R}^2$ unless (positive) $x$ is restricted to half of the plane, with negative values extending to the other half. The curve $y=ax$ in this system is a skewed conic section (parabola for $a=1$, ellipse for $a>1$, or hyperbola for $0<a<1$).

\textbf{Third Manner:} $x$ and $y$ each extend radially from separate given points. In modern parlance these are ``two-center bipolar coordinates'' (not to be confused with ``bipolar coordinates,'' which use the quantity $\ln(x/y)$ together with the angle between $x$ and $y$). The system finds applications in problems involving electric dipoles, two-source interference, and orbital motion. The curve $y=ax$ in this system is an Apollonian circle (or a straight vertical line if $a=1$).

\textbf{Fourth Manner:} $x$ and $y$ both extend radially from the same given point, but $x$ only extends to a given horizontal axis. This system is related to polar coordinates by the transformation $x=y_0\csc(\theta)$, where $y_0$ is the vertical distance from the radial point to the axis. Newton gives no examples for this and the following two manners, but one may note that the curve $y=ax$ in this system is a horizontal line.

\textbf{Fifth Manner:} Newton uses the same figure for this manner as the previous and subsequent manners, though they lead to slightly different coordinate charts. Here $y$ is still the radius from a given point, but $x$ is now the distance along the horizontal axis (but not an abscissa) from a given point to the radius-axis point of intersection. This system is related to polar coordinates by the transformation $x=x_0+y_0\tan(\theta)$, where $x_0$ and $y_0$ are the horizontal and vertical distances between the two given points. The curve $y=ax$ in this system is the conchoid of a kappa curve.

\textbf{Sixth Manner:} Instead of defining $y$ as the radius as in the previous two manners, $y$ is now the distance along the radius from the axis to the coordinate point. This system has no known modern equivalent, but note that the constant-$y$ contours form conchoids of Nicomedes. The curve $y=ax$ in this system has no straightforward classification but can be expressed in polar coordinates as the linear combination of $\tan(\theta)$, $\csc(\theta)$, and a constant.

\textbf{Seventh Manner:} $x$ measures the distance along the circumference of a given circle to the coordinate point's radial axis, while $y$ measures the radius. This system is related to polar coordinates by the transformation $x=R\theta$, where $R$ is the radius of the given circle. The curve $y=ax$ in this system gives an Archimedean spiral.

\textbf{Eighth Manner:} This semi-polar system has $x$ defined as in the seventh manner, while $y$ is a standard Cartesian ordinate. The curve $y=ax$ in this system is called by Newton the ``Quadratrix of the Ancients'' (it was first used by Hippias and Dinostratus to square the circle), which can be created by the rolling shutter effect when viewing rotating objects.

\textbf{Ninth Manner:} $x$ measures the distance along an arbitrary given curve, and $y$ measures the distance from this curve along a line parallel to a given skewed axis. Though not shown here, this figure is Newton's only instance where he explicitly shows a coordinate ($y$) running in either direction (above or below the given curve). This system is the most general set of curvilinear coordinates he considers. Newton mentions that if the given curve is a circle, then the curve $y=ax$ gives a trochoid (specifically a cycloid when $a=1$), which has seen numerous applications in physics, \textit{e.g.} the brachistrochrone and tautochrone problems and the path of a charged particle in crossed electric and magnetic fields.

Newton also mentions two additional coordinate systems that can be derived from that of Fig.~\ref{fig:coord}i, which are different from that coordinate grid only in the spacing of the constant-$x$ contours: if $y$ is kept the same, $x$ can be changed either to denote the dashed line or the abscissa (the distance from the origin to the lower endpoint of the dashed line).

% \begin{acknowledgments}
% \end{acknowledgments}

\section{Author declarations}
The authors have no conflicts of interest to disclose.

\end{document}